\documentstyle[12pt]{article}

\begin{document}

\baselineskip=21pt

\centerline{\bf Experimental Realization of Teleporting an Unknown}
\centerline{\bf Pure Quantum State via Dual Classical and 
Einstein-Podolski-Rosen Channels}
\vskip 6mm

\centerline{D. Boschi$^{(1)}$, S. Branca$^{(1)}$, F. De
Martini$^{(1)}$,}
\centerline{L. Hardy$^{(1)}$, S. Popescu$^{(2, 3)}$}

\vskip 6mm

\centerline{\it ${}^{(1)}$Dipartimento di Fisica, Instituto Nazionale
di Fisica Nucleare,}
\centerline{\it Instituto Nazionale di Fisica della Materia, Universit\`a
\lq\lq La Sapienza",}
\centerline{\it Roma 0018, Italy}

\vskip 3mm

\centerline{\it ${}^{(2)}$Isaac Newton Institute, University of Cambridge,}
\centerline{\it Cambridge CB3 0EH, UK}

\vskip 3mm
\centerline{\it ${}^3$BRIMS, Hewlett-Packard Labs.,}
\centerline{Bristol BS12 SQZ, U.K.}

\vskip 6mm

We report on a quantum optical experimental implementation of teleportation of
unknown pure quantum states.  This realizes all the nonlocal aspects of the
original scheme proposed by Bennett et al.  and is equivalent to it up to a
local operation.  We exhibit results for the teleportation of a linearly
polarised state and of an elliptically polarised state.  We show that the
experimental results cannot be explained in terms of a classical channel alone.

\vskip 6mm


\vskip 10mm

In [1], Bennett et al showed that  an unknown
quantum state can be  \lq\lq  disassembled into, then later
reconstructed from,
purely classical information and purely nonclassical
Einstein-Podolsky-Rosen (EPR) correlations."  They called this process
{\it teleportation}.
In their scheme, a sender, traditionally called Alice, is given a state
unknown to her.  She also has one of two particles prepared in an EPR
state (such as a singlet state).  She performs a Bell measurement on
the combined system of the unknown state and her EPR particle
and transmits the result of this measurement by a classical channel
to Bob who has the second of the EPR particles.
Depending on the result of the measurement, Bob performs one of four
possible unitary transformations on his particle and it will now be in
the unknown  state.

In the experiment reported in this paper we take an approach first
suggested in [2] in which a total of two photons, rather
than three, are used.  The EPR state is realized by
${\bf k}$-vector (or path) entanglement and the polarisation degree of freedom
of one of the photons is employed for preparing the unknown state.
This avoids the difficulties associated with having three photons and,
as will be seen below,
makes the Bell measurement much more straight forward (indeed, in three
photon schemes, where the Bell measurement is made on two photons,
feasible schemes can have only a 50\% rate of success even in the ideal case
[3]).  However, this approach does place a restriction on us in that the
preparer must prepare his state on one of the EPR photons, and so the
unknown state cannot come from outside.  Nevertheless, the scheme to
be described here realizes all the nonlocal aspects of the original
teleportation scheme, and is equivalent to the original scheme upto a
local operation (since, in principle, the unknown state of a particle
from outside could be swapped onto the polarisation degree of freedom of
Alice's EPR particle by a local unitary operation).
In particular, as in the original scheme, we emphasize
that if the preparer does not tell Alice what state he has prepared then
there is no way Alice can find out what the state is.

No experiment is perfect, so we need to know
how good the experiment has to be before we can say we have quantum
teleportation.  We will take as our objective to show that
the experimental results cannot be explained by a classical channel
alone (that is without an EPR pair).
Thus, consider the following
scenario.  With probability of ${1\over3}$ the preparer prepares one of
the states $|\phi_a\rangle$ ($a=1,2,3$) which correspond to equally
spaced linearly polarised states at $0^\circ$, $120^\circ$,
$-120^\circ$.
He then gives this state to Alice (without telling her which one it is).
Alice makes
a measurement on it in an attempt to gain some information about the
state.  The most general measurement she can make is a positive operator
valued measure.  She will never obtain more information if some of the
positive operators are not of first rank, and thus we can take them all
to be of first rank, that is proportional to projection operators.  Let Alice's
measurement have $L$ outcomes labeled $l=1,2,\dots L$ and let the positive
operator associated with outcome $l$ be
$|\varepsilon_l\rangle\langle\varepsilon_l|$.  We require that
\begin{equation}\label{povm}
\sum_{l=1}^L |\varepsilon_l\rangle\langle\varepsilon_l| = I
\end{equation}
Note that in general the states $|\varepsilon_l\rangle$ are neither
orthogonal to each other or normalized but rather form an overcomplete
basis set.
The probability of getting outcome $l$ given that the state prepared is
$|\phi_a\rangle$ is $|\langle\phi_a|\varepsilon_l\rangle|^2$
Alice sends  the information $l$ to Bob over the classical channel and Bob
prepares a state $|\phi^{c}_l\rangle$  (the $c$ denotes that the
state has been \lq\lq classically teleported").  This state is chosen so as
to give the best chance of passing a test for the original state.  Bob
now passes this state onto a verifier.
We suppose that the preparer has told the verifier
which state he prepared and the verifier sets his apparatus to measure
the projection operator, $|\phi_a\rangle\langle\phi_a|$, onto this state.
The probability that the
classically teleported state will pass the test in this case is
$|\langle\phi_a|\phi^c_l\rangle|^2$
The average probability, $S$, of passing the test is given by
\begin{equation}\label{sprob}
S= \sum_{a,l}  {1\over3}|\langle\phi_a|\phi^c_l\rangle|^2
|\langle\phi_a|\varepsilon_l\rangle|^2
\end{equation}
In the appendix we show this classical teleportation protocol must
satisfy
\begin{equation} \label{ctl}
S\leq {3\over4}
\end{equation}
To show that we have quantum teleportation we must show that the
experimental results violate this inequality [4].

The experiment is shown in fig. 1.  Pairs of polarisation entangled photons
are created directly using type II degenerate parametric downconversion by the
method described in [5,6].  The crystal is pumped by a 200mW UV cw argon
laser with wavelength
351.1nm.  The downconverted photons have wavelength 702.2nm.  The state
of the photons at this stage is
${1\over\sqrt{2}}(|v\rangle_1|h\rangle_2+|h\rangle_1|v\rangle_2)$
However, we want a {\bf k}-vector entangled state so next we let each
photon pass through a calcite crystal after which the state becomes
\begin{equation}
{1\over\sqrt{2}}(|a_1\rangle|a_2\rangle+|b_1\rangle|b_2\rangle)
|v\rangle_1|h\rangle_2
\end{equation}
By this method a polarisation entangled state has been converted into a
{\bf k}-vector entangled state.
Here, $|a_1\rangle|v\rangle_1$, for example, represents the state of
photon 1 in path $a_1$ and having vertical polarisation.  Since each
photon has the same polarisation in each of the two paths it can take,
the polarisation part of the state factors out of the {\bf k}-vector
entanglement.   The EPR pair for the teleportation procedure is
provided by this {\bf k}-vector
entanglement.  By means of quarter wave plates orientated at some angle
$\gamma$  to the vertical and Fresnel romb polarisation rotators
acting in the same way on paths $a_1$ and $b_1$ as shown in Fig. 1 the
polarisation degree of freedom of photon 1 is
used by the preparer to prepare the state
$|\phi\rangle=\alpha|v\rangle_1+\beta|h\rangle_1$.  This is the state
to be teleported.  The state of the whole system is now
\begin{equation}\label{pretele}
{1\over\sqrt{2}}(|a_1\rangle|a_2\rangle+|b_1\rangle|b_2\rangle)
(\alpha|v\rangle_1+\beta|h\rangle_1)|h\rangle_2
\end{equation}
We now introduce four orthonormal states which are directly analogous
to the Bell states considered in [1]:
\begin{equation}\label{cpm}
|c_{\pm}\rangle={1\over\sqrt{2}}(|a_1\rangle|v\rangle_1\pm |b_1\rangle
|h\rangle_1)
\end{equation}
\begin{equation}\label{dpm}
|d_{\pm}\rangle={1\over\sqrt{2}}(|a_1\rangle|h\rangle_1\pm |b_1\rangle
|v\rangle_1)
\end{equation}
We can rewrite (\ref{pretele}) using these states as a basis:
\begin{eqnarray}\label{jbm}
{1\over2}|c_+\rangle(\alpha|a_2\rangle+\beta|b_2\rangle)|h\rangle_2
+{1\over2}|c_-\rangle(\alpha|a_2\rangle-\beta|b_2\rangle)|h\rangle_2
\nonumber \\
+{1\over2}|d_+\rangle(\beta|a_2\rangle+\alpha|b_2\rangle)|h\rangle_2
+{1\over2}|d_-\rangle(\beta|a_2\rangle-\alpha|b_2\rangle)|h\rangle_2
\end{eqnarray}
For Alice, it is simply a question of measuring on the basis $|c_\pm\rangle$,
$|d_\pm\rangle$. To do
this we first rotate the polarisation of path $b_1$ by a further
$90^\circ$ (in the actual experiment this was done by setting the angle
of the Fresnel romb in path $b_1$ at $\theta+90^\circ$ rather than by
using a separate plate as shown in the figure) so that the state
$|b_1\rangle|v\rangle_1$ becomes
$-|b_1\rangle|h\rangle_1$ and the state $|b_1\rangle|h\rangle_1$ becomes
$|b_1\rangle|v\rangle_1$. Thus,
\begin{equation}
|c_{\pm}\rangle \rightarrow {1\over\sqrt{2}}
(|a_1\rangle\pm |b_1\rangle)|v\rangle_1)
\end{equation}
\begin{equation}
|d_{\pm}\rangle\rightarrow{1\over\sqrt{2}}
(|a_1\rangle\mp |b_1\rangle)|h\rangle_1)
\end{equation}
Paths $a_1$ and $b_1$ now impinge on the two
input ports of an ordinary 50:50 beamsplitter ($BS$).
At this beamsplitter each of the two polarisations $h$, and $v$
interfere independently.  After the beamsplitter there are two
polarisers which are set either to transmit $h$ or to transmit $v$
polarisation.  When the path lengths have been set appropriately, the state
${1\over\sqrt{2}}(|a_1\rangle\pm |b_1\rangle)$ gives rise to a click at
detector $D_{A_\pm}$.  If the polarisers are set to $v$ then a click at
$D_{A_\pm}$ corresponds to measurement onto $|c_{\pm}\rangle$ and if the
polarisers are set to $h$ then a click at $D_{A_\pm}$ corresponds to a
measurement onto $|d_\mp\rangle$.  In this way each of the four Bell
states can be measured.

At Bob's end, path $b_2$ is rotated through $90^\circ$ by a half wave
plate.   Then paths $a_2$ and $b_2$ are combined at a polarising
beamsplitter orientated to transmit vertical and reflect horizontal
polarisation.  The state in (\ref{jbm}) becomes
\begin{eqnarray}\label{pjbm}
{1\over2}|c_+\rangle(\beta|v\rangle_2+\alpha|h\rangle_2 )
+{1\over2}|c_-\rangle(-\beta|v\rangle_2+\alpha|h\rangle_2)
\nonumber \\
+{1\over2}|d_+\rangle(\alpha|v\rangle_2+\beta|h\rangle_2)
+{1\over2}|d_-\rangle(-\alpha|v\rangle_2+\beta|h\rangle_2)
\end{eqnarray}
Bob's photon can be transformed back to the original state
$\alpha|v\rangle+\beta|h\rangle$ by applying an appropriate unitary
transformation depending on the outcome of Alice's measurement.
However, we are simply interested in verifying that the appropriate state
has been produced at Bob's end so rather than performing these
unitary transformations we will simply orientate the measuring apparatus
at end 2 appropriately for each of Alice's outcomes (the transformations
can either be seen as active transformations or as a passive
re-orientation of our reference system with respect to which
verification measurements are made).
In the case of linear polarisation the verification measurements can be
accomplished by rotating the polarisation of the state through an an
angle $\theta_B$ by means of Fresnel romb devices,
then letting it impinge on a
polarising beamsplitter followed by two detectors $D_B(\theta_B)$ (a photon
originally incident with polarisation $\theta_B$ would certainly be
detected at this detector), and $D_B(\theta_B^\perp)$. In the more general
case of elliptical polarisation we can add a quarter wave plate orientated at
an angle $\gamma_B$ to the vertical before the polarisation rotator.

Wide filters ($\Delta\lambda$=20nm) were placed just before each detector.
Wide, rather
than narrow, filters were used so that the count rate was high enough to
allow measurements to be made in a few seconds in order that problems
associated with phase drift were minimize.

The beamsplitter $BS$ and the trombone arrangement in path $a_2$ were
each mounted on a computer controlled micrometrical stage which could be
incremented in 0.1$\mu$m steps.  To align the system a half wave plate
orientated at $45^\circ$ to the vertical was
placed before one of the calcite crystals to rotate the polarisation by
$90^\circ$.  This had the effect of sending both photons to Alice's end
or Bob's end.  The correct values of $\Delta z$ and $\Delta y$ were
found by looking for an interference dip in the coincidence count rates
between $A_+$ and $A_-$ at Alice's end, and between $D_B(45^\circ)$ and
$D_B(-45^\circ)$ at Bob's end.  After alignment, the half wave plate
was rotated to $0^\circ$  to the vertical so that it had no effect on
the polarisation of photons passing through it.

We will report separately two aspects of the experiment.
First using three equally spaced
linear polarisation settings ($0^\circ$, $+120^\circ$, $-120^\circ$)
we will see that the classical
teleportation limit in equation (\ref{ctl}) is surpassed.  Second we
will see that, for some arbitrary states (linear and elliptically
polarised), we see all the expected features.

Let $I(\theta_B)$ be the coincidence count between $D_B(\theta)$ and one of
Alice's detectors.  Let $I_\parallel$ be the coincidence rate when
$\theta_B$ is orientated so as to measure the projection operator onto
the corresponding term in (\ref{pjbm}).  For example, if $\theta=120^\circ$
then
Alice's output $|c_+\rangle$ corresponds requires $\theta_B=-30^\circ$.  Let
$I_\perp$ be the count rate when $\theta_B$ is rotated through
90$^\circ$ from the value used to measure $I_\parallel$.
We will have $I_\parallel = k |\langle\phi|\phi_{\rm tele}\rangle|^2$
and $I_\perp = k |\langle\phi|\phi^\perp_{\rm tele}\rangle|^2$
where $|\phi\rangle$ is the prepared state and $|\phi_{\rm tele}\rangle$
is the state that is actually produced at Bob's side by the teleportation
process.  If the state produced at Bob's end is not pure then this
analysis is easily adapted by summing over a particular decomposition of
the impure state.  The normalization constant $k$ depends on the detector
efficiencies.  Since $I_\parallel + I_\perp=k$, we have that
\begin{equation}\label{fid}
|\langle\phi|\phi_{\rm tele}\rangle|^2  = {I_\parallel\over
I_\parallel+I_\perp}
\end{equation}
To beat the classical teleportation limit the average value of this
quantity must exceed $3/4$.  This average is taken over the three
equally spaced linear polarisations (each weighted with probability
${1\over3}$) and over each of the four possible
outcomes of Alice's Bell state measurement (each weighted by ${1\over4}$).
With the average understood to be in this sense we can write
\begin{equation}
S=\Big[{I_\parallel\over I_\parallel+I_\perp} \Big]_{\rm av}
\end{equation}
This quantity was measured and we found that
\[ S=0.831 \pm 0.009 \]
This represents a violation of the classical teleportation limit by 9
standard deviations.   Note, if $I_\parallel=I_{\rm max}$ and
$I_\perp=I_{\rm min}$ (as we would expect, and as is indeed true for the
data to be discussed below), then if the quantity in (\ref{fid})
greater than $3/4$ then the visibility is greater than 50\% (and vice
versa).

Now consider in more detail one linear polarisation case.
All the quarter wave plates were removed in this case.
In Fig. 2, we show coincidence count rates between each outcome of
Alice's Bell state measurement ($c_{\pm}$ and $d_{\pm}$) and Bob's detector
$D_B(\theta_B)$ as a function of $\theta_B$ for the particular case where
the preparer prepared linear polarisation with $\theta=22.5^\circ$ with
respect to the horizontal direction.
Note that the displacements of the maxima are 22.5$^\circ$, 67.5$^\circ$,
112.5$^\circ$ and -22.5$^\circ$.  These are  consistent with equation
(\ref{pjbm}) where $\alpha=\sin(\theta)$ and $\beta=\cos(\theta)$.

To prepare
an elliptical polarised state we set $\theta=0$ and inserted the quarter
wave plates at angle $\gamma$ equal to  $20^\circ$ to the vertical.
This produced the elliptically polarised state
\[
{1\over\sqrt{2}}[(1+i\cos(2\gamma))|\updownarrow\rangle
+\sin(2\gamma)|\leftrightarrow\rangle]
\]
To verify that the state had been teleported according to equation
(\ref{pjbm}) quarter wave plate was used at Bob's end orientated at
angle $\gamma_B$.  A different setting of this was used corresponding to
each of Alice's outcomes.  For outcomes $|d_{\pm}\rangle$ we set
$\gamma_B=\pm\gamma+90^\circ$.  This converts the corresponding state
at Bob's side to $|v\rangle$.  For outcomes $|c_{\pm}\rangle$ we set
$\gamma_B=\pm \gamma$.  This converts the corresponding state at Bob's
side to $|h\rangle$.  Then the state was analysed in linear polarisation
over a range of values of $\theta_B$.  The results, shown in fig. 3,
demonstrate that the state after the quarter wave plate is vertically or
horizontally polarised as required.

In this paper we have seen how a state which is unknown to Alice can
be disassembled into purely classical and purely nonlocal EPR
correlations, and then it can be reconstructed at a distant location.
In the reconstruction procedure we took an essentially passive view
of the unitary transformations.  That is to say the verifier took them
into account in his verification process by orientating his apparatus
appropriately.  Work is currently under progress to implement the
transformations in an active way by using fast Pockel cells fired by
Alice's detectors. Also, work is under way to use polarising
beamsplitters rather than the polarisers at Alice's end, followed by
four detectors.  This would allow an \lq in principle' teleportation
efficiency  of 100\%.

{\bf Appendix}

In this appendix we prove inequality (\ref{ctl}) for the classical
teleportation protocol.  Define the normalized state
$|\Omega_l\rangle$ by
$|\Omega_l\rangle=\sqrt{\mu_l}|\varepsilon_l\rangle$ where
$\mu_l=\langle\varepsilon_l|\varepsilon_l\rangle$.
By taking the trace of (\ref{povm}) we obtain
\begin{equation}\label{smu} \sum_l \mu_l=2 \end{equation}
The 2 here corresponds to the dimension of the Hilbert space.
Define
\begin{equation}\label{tl}
T_l=\sum_a |\langle\phi_a|\phi^c_l\rangle|^2
|\langle\phi_a|\Omega_l\rangle|^2
\end{equation}
Then
\begin{equation}
S=\sum_l {1\over3}\mu_l T_l
\end{equation}
By varying with respect to the vectors $|\phi_l^c\rangle$ and
$|\Omega_l\rangle$ we obtain $T^{\rm max}_l={9\over8}$.
Hence,
\[ S\leq  \sum_l {1\over3}\mu_l{9\over8} \]
By using (\ref{smu}) we obtain (\ref{ctl}) as required.

{\bf Acknowledgements}

We would like to thank C. Bennett, H. Bernstein,
G. Di Giuseppe, D. Di Vincenzo and H. Weinfurter for discussions, and
the CEE-TMR (contract number ERBFMRXCT96-0066) and INFM (contract
PRA97-cat) for funding.

{\bf References}

[1]  C. Bennett, G. Brassard, C. Crepeau, R. Jozsa, A. Peres and W.
Wootters, Phys. Rev. Lett. {\bf 70}, 1895 (1993).

[2] S. Popescu, {\it An optical method for teleportation}, preprint.

[3] H. Weinfurter, Europhys. Lett. 559 (1994); S. L. Braunstein and A.
Mann, Phys. Rev. A {\bf 51}, R1727 (19955); K. Mattle, H. weinfurter,
P. G. Kwiat and A. Zeilinger, PRL {\bf 76}, 4656 (1996).

[4] By taking a uniform distribution of states over the Poincare sphere,
a lower upper bound is obtained in
S. Massar and S. Popescu, Phys. Rev. Lett. {\bf 74}, 1259 (1995).

[5] D. Klyshko, {\it Photons and Nonlinear Optics} (Gordon and Breach,
New York, 1988).

[6] P. G. Kwiat, K. Mattle, H. Weinfurter, A. Zeilinger, A. V. Sergienko
and Y. H. Shih, Phys. Rev. Lett. {\bf 75}, 4337 (1995); A. Garuccio, in
{\it Fundamental Problems in Quantum Theory: A Conference Held in Honor
of Professor John A. Wheeler} (Annals of the New York Academy of
Sciences, Vol. 755, 1995); B. Boschi, PhD Thesis, Universit\`a di Roma
\lq\lq La Sapienza", July 1995).

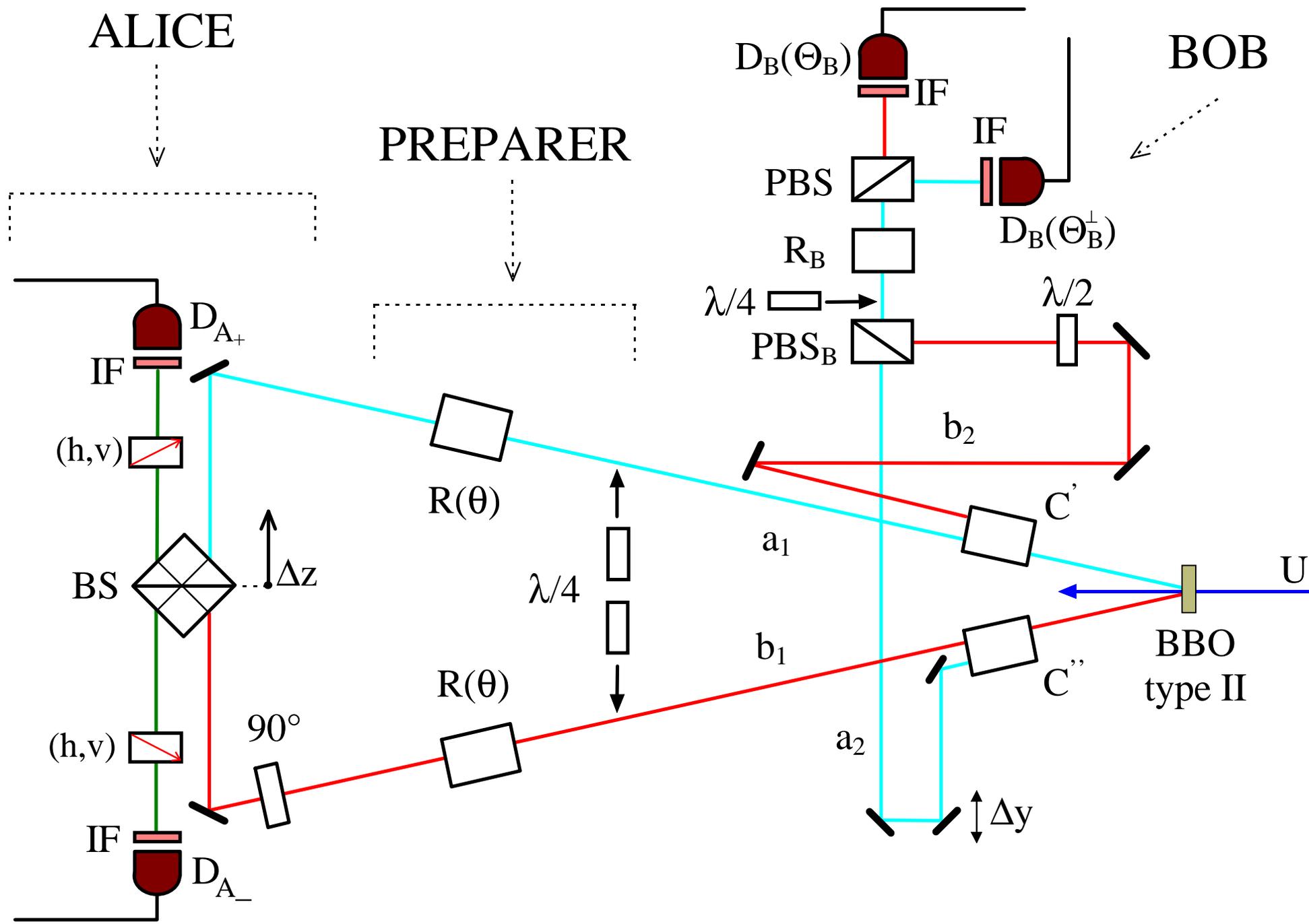
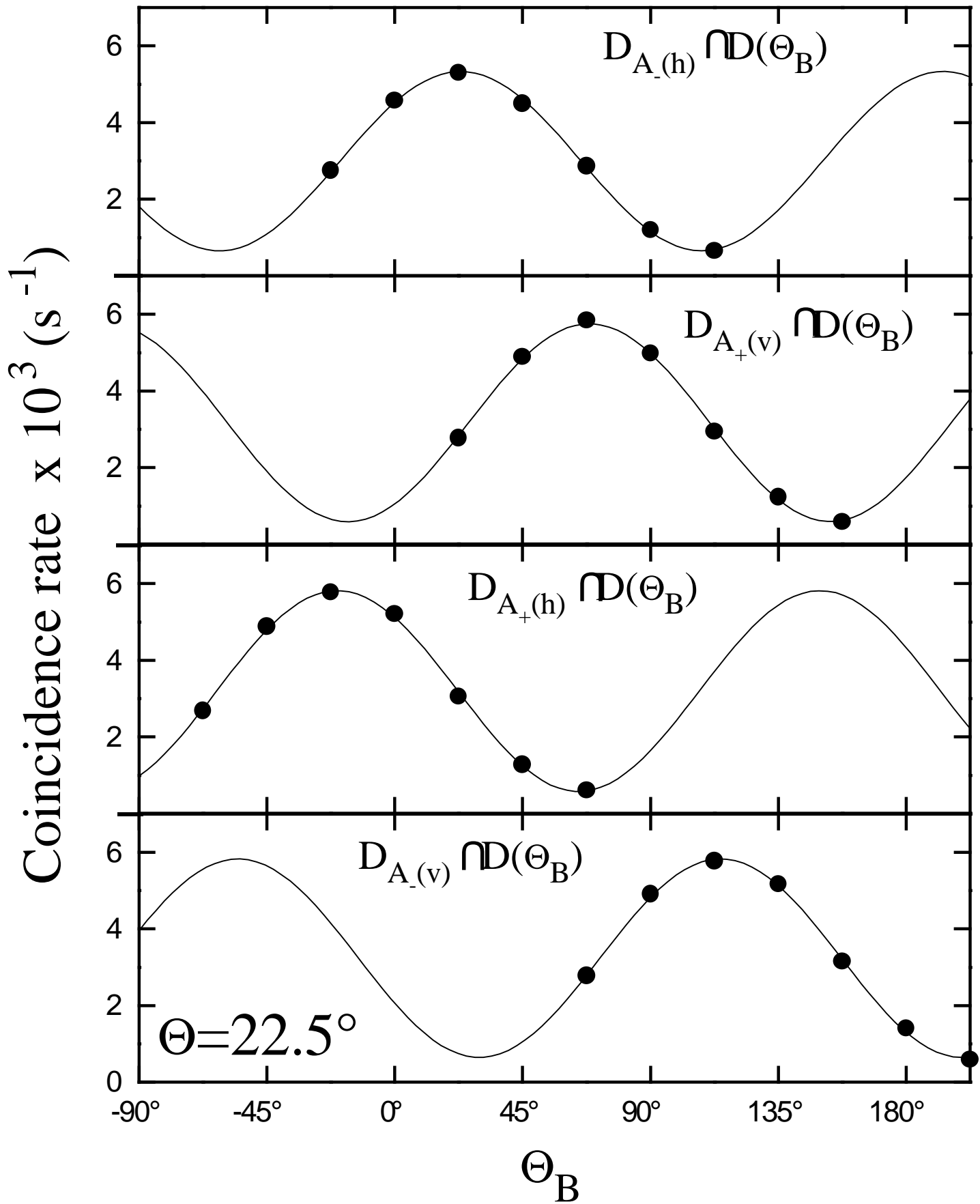
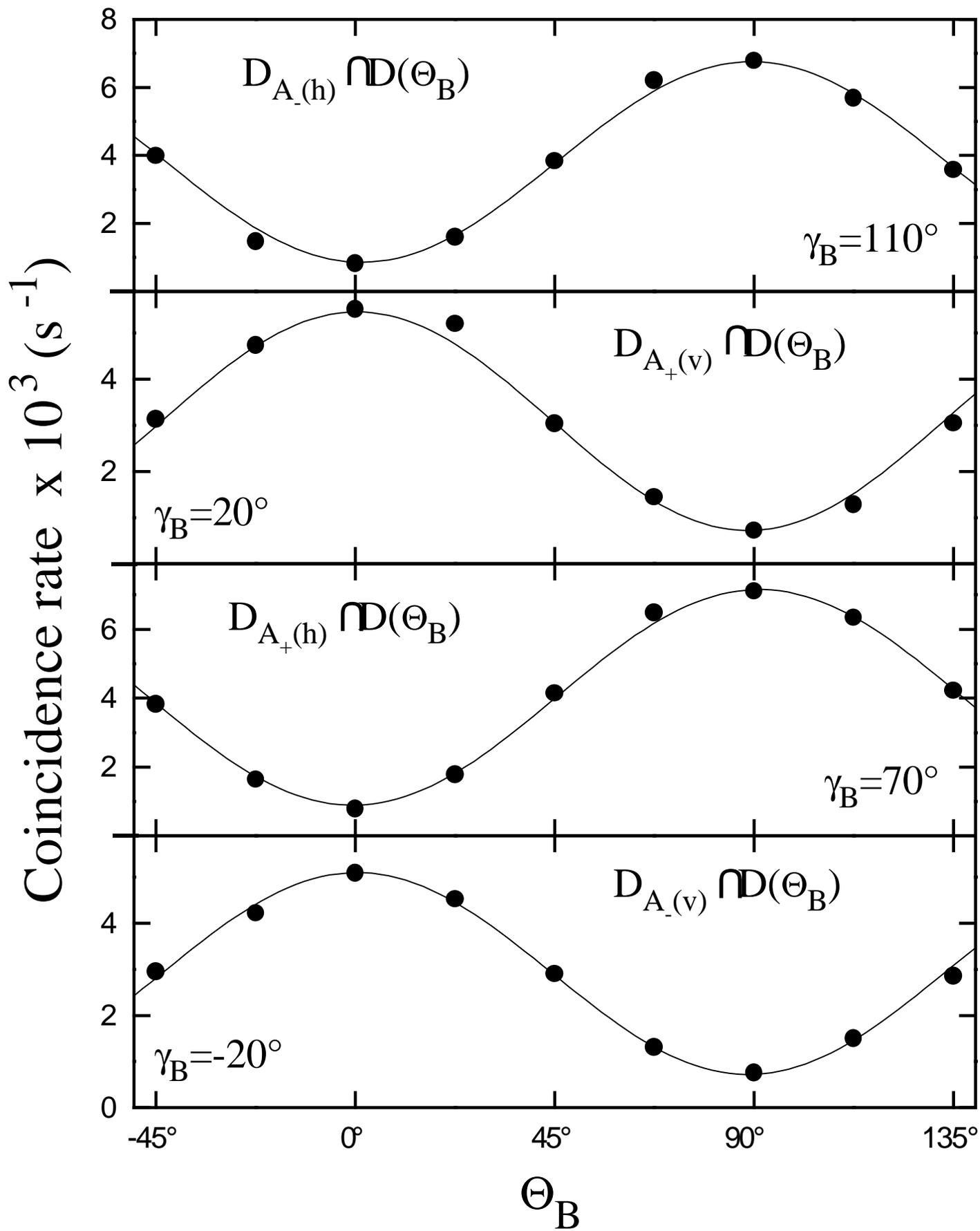
\end{document}